\pdfoutput=1

\documentclass[RNAAS]{aastex62}
\usepackage{graphicx,epsf,soul,xcolor,gensymb}

\newcommand{\scc}{\texttt{shock\_cooling\_curve }}

\begin{document}

\title{\texttt{shock\_cooling\_curve}: A Python-Based Package for Extensive and Efficient Modeling of Shock Cooling Emission in Supernovae}

\correspondingauthor{Padma Venkatraman}
\email{vpadma@stanford.edu}

\author[0000-0001-8638-2780]{Padma Venkatraman}
\affiliation{Kavli Institute for Particle Astrophysics and Cosmology, Department of Physics, Stanford University, Stanford, CA 94309, USA}
\affiliation{Department of Astronomy, University of California, Berkeley, CA 94720-3411, USA}

\author[0000-0002-3934-2644]{Wynn Jacobson-Gal\'an}
\affiliation{Department of Astronomy, University of California, Berkeley, CA 94720-3411, USA}


\begin{abstract}
The light-curve evolution of a supernova contains information on the exploding star. Early-time photometry of a variety of explosive transients, including Calcium-rich transients and type IIb/Ibc and IIP supernovae shows evidence for an early light curve peak as a result of the explosion's shock wave passing through extended material (i.e., shock cooling emission (SCE)). Analytic modeling of the shock cooling emission allows us to estimate progenitor properties such as the radius and mass of extended material (e.g., the stellar envelope) as well as the shock velocity. In this work, we present a Python-based open-source code that implements four analytic models originally developed in \citealt{p15}, \citealt{p20} and \citealt{sw17} applied to photometric data to obtain progenitor parameter properties via different modeling techniques (including non-linear optimization, MCMC sampling). Our software is easily extendable to other analytic models for SCE and different methods of parameter estimation. 
\end{abstract}

\section{Introduction}

A shock wave associated with a stellar explosion initiated close to the core of a supernova (SN) progenitor will propagate outwards, heating up optically thick stellar material as it proceeds. When the density of the material in front of the shock decreases, and its optical depth drops below a critical value of $\tau \sim c/v_{sh}$, where $v_{sh}$ refers to the shock velocity, radiation can leave the region and reach the observer for a time-span of hours - this is referred to as ``shock breakout'' \citep{sbt}. This forms the first light curve peak - we typically miss the shock breakout signal due to its short time period. After this, the previously shock heated extended material cools down and the stored energy is released as electromagnetic radiation, which we refer to as ``shock cooling emission,'' (SCE) forming what is typically the first observed peak in an SN light curve (followed by the Nickel-powered peak that characterizes H-stripped SN when there is no other energy source).

\vspace{2mm}
Modeling the cooling-envelope emission enables constraints on the different key unknown parameters of a SN's progenitor system. Early-time photometry of Calcium-rich (e.g., \citealt{caRichZ, 2019ehk, jg22}) transients, type IIb (e.g., \citealt{arcavi, carich11bz}), Ibc (e.g., \citealt{IbcZ}), and type IIP (e.g., \citealt{Hosseinzadeh18, Morag23}) supernovae (SNe) show evidence for a primary light curve peak that results from SCE.
\begin{figure*}
    \centering
    \includegraphics[width=\textwidth]{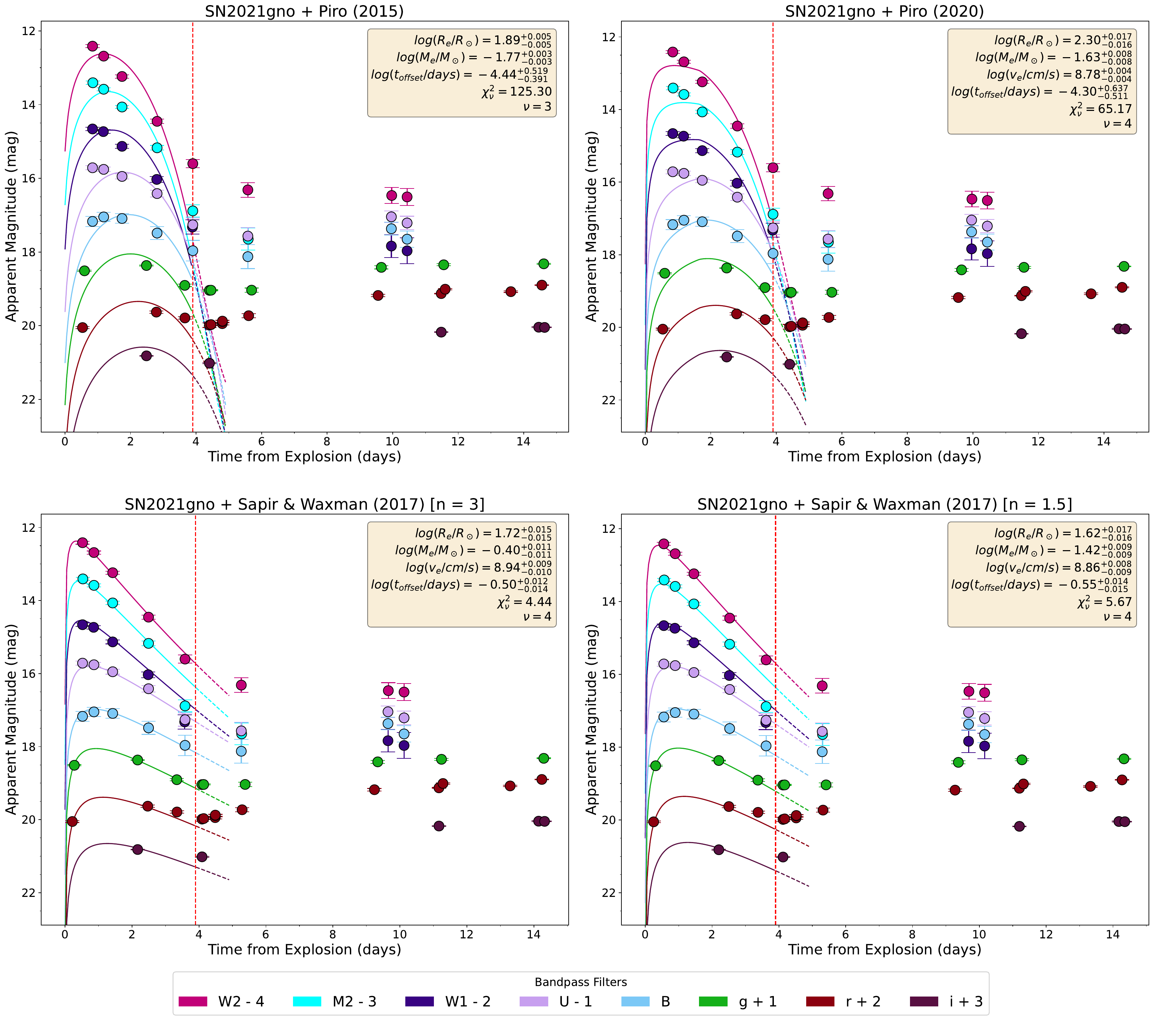}
    \caption{Best-fit results with 1-$\sigma$ errors for SN2021gno obtained from MCMC sampling. The observed photometry is presented in AB magnitude system \citep{jg22}. Vertical red dashed line marks end of SCE timescale. Best-fit parameters along with the goodness-of-fit reduced chi-squared ($\chi^2_{\nu}$) value is provided at the top right. Filters are shown at the bottom of the figure.}
    \label{fig:oneplot}
\end{figure*}
\vspace{2mm}
We present a open-source software package \scc actively being developed on \texttt{GitHub}\footnote{\url{https://github.com/padma18-vb/shock\_cooling\_curve}}, packaged via \texttt{PyPI}.
This package derives key progenitor properties (extended material radius/mass, shock velocity and time offset) from the following analytic SCE models: P15 \citep{p15}, P20 \citep{p20}, SW17 \citep{sw17}. This tool can assist in efficient modeling and analysis of light curves spanning several SN sub-types, using models incorporating different physical assumptions, thus probing the systematic uncertainties behind each model.

\section{Software}\label{soft}



\scc requires photometric data (.csv) and configuration (.ini) files. The tool recognizes data in SDSS and Johnson bandpass filters with a conversion available between AB and Vega magnitude systems. The data file should contain photometric magnitudes, filters, dates, and a binary value specifying magnitude system. The configuration file should contain SN information e.g., SCE timescale, reddening, core mass, kinetic energy, opacity, and distance.

\vspace{2mm}
\scc fits the supplied photometry to a specific SCE model and outputs best-fit key progenitor parameters. 
The models output a temperature and luminosity assuming blackbody conditions using uniformly sampled parameter inputs. \scc uses \texttt{pysynphot} \citep{pysynphot} to create synthetic photometry by convolving the observing instrument's bandpass with a blackbody spectrum generated by the SCE models. This synthetic photometry is repeatedly generated for a range of SCE parameters and the best-fit parameters minimize the log-likelihood between synthetic and observed photometry. 


\vspace{2mm}




\vspace{2mm}

\scc has three modeling methods: (1) \texttt{nl\_curve\_fit} adopts non-linear least squares modeling (2) \texttt{minimize} adopts minimization techniques for analytical models with discontinuous luminosity functions and (3) \texttt{MCMC\_fit} applies Markov Chain Monte Carlo (MCMC) sampling (using \texttt{emcee} \citep{emcee}, adopting parallelization techniques). The fitting module stores best-fit parameters and uncertainty margins for different fitting methods, MCMC sampler chains and goodness-of-fit estimates. The plotting module plots posterior parameter corner plots, MCMC sampler chains and best-fit light curves (shown in Fig. \ref{fig:oneplot}) and parameter-varying light curve movies.

\vspace{2mm}


\section{Discussion and Conclusions}\label{dc}
To demonstrate this tool, we re-model SN~2021gno \citep{jg22, Ertini23} and present the results in Figure \ref{fig:oneplot}. To test the performance of our software, we model type IIb SN~2016gkg and compare the best-fit parameters to \cite{arcavi}. Overall, our fitting results obtained by using the P15 and SW17 models are statistically consistent within $1\sigma$ of the results presented in \cite{arcavi}. We note our application of the SW17 (n=3) model produces better fits (based on reduced chi-squared metrics), but returns an unphysical shock velocity - this could be due to constraints placed by the analytic formalism. 

As shown in Figure \ref{fig:oneplot}, the choice of analytic model can yield a variety of best-fit parameters of the extended material. Modeling the SCE using all available models can allow us to obtain constraints on the progenitor properties, and quantitatively assess the assumptions made in each fit. Applying these models to early-time data allows us to effectively rule out or posit that the primary peak in double-peaked observations is a result of SCE. This package’s modular nature also allows for inclusion of more recently developed analytical models (e.g., \citealt{Morag23}).

\vspace{2mm}

We thank Raffaella Margutti for valuable discussions and comments on the manuscript.
\bibliography{SCC}
\end{document}